```
Title: (tn_header.eps)
Creator: Adobe Illustrator(r) 6.0
CreationDate: (17/04/97) (16:46)
```

May 31, 1998

# Detector Construction Management and Quality Control: Establishing and Using a CRISTAL System

J.-M. Le Goff

*CERN, Geneva, Switzerland*

G. Chevenier[1]

*Institut fur Teilchenphysik, Eidgenossische Technische Hochschule, Zurich, Switzerland*

A. Bazan, T. Le Flour, S. Lieunard, S. Murray, J.-P. Vialle

*LAPP, IN2P3, Annecy-le-Vieux, France*

N. Baker, F. Estrella[1], Z. Kovacs[1], R. McClatchey[1]

*Dept. of Computing, Univ. West of England, Frenchay, Bristol BS16 1QY UK*

G. Organtini

*Universita di Roma e Sezione dell' INFN, Roma, Italy*

S. Bityukov

*IHEP, Protvino, Russia*

### Abstract

The CRISTAL (Cooperating Repositories and an Information System for Tracking Assembly Lifecycles) project is delivering a software system to facilitate the management of the engineering data collected at each stage of production of CMS. CRISTAL captures all the physical characteristics of CMS components as each sub-detector is tested and assembled. These data are retained for later use in areas such as detector slow control, calibration and maintenance. CRISTAL must, therefore, support different views onto its data dependent on the role of the user. These data viewpoints are investigated in this paper. In the recent past two CMS Notes have been written about CRISTAL. The first note, CMS 1996/003, detailed the requirements for CRISTAL, its relationship to other CMS software, its objectives and reviewed the technology on which it would be based. CMS 1997/104 explained some important design concepts on which CRISTAL is and showed how CRISTAL integrated the domains of product data management and workflow management. This note explains, through the use of diagrams, how CRISTAL can be established for detector production and used as the information source for analyses, such as calibration and slow controls, carried out by physicists. The reader should consult the earlier CMS Notes and conference papers for technical detail on CRISTAL – this note concentrates on issues surrounding the practical use of the CRISTAL software.

---

[1] Now at CERN

# 1. Background

Each CMS detector will be constructed out of a large number of precision parts and will be produced and assembled during the next decade by centres distributed worldwide. Each constituent part of each detector must be measured and tested locally prior to its assembly and integration in the experimental area at CERN. The Barrel ECAL detector, for example, will have its crystals grown and part-characterised in SIC (China) and Bogorodisk (Russia), its APDs assembled and tested at IPN (Lyon) and its alveoli structures supplied by LPNHE (see figure 1). Detector elements constitute parts of ECAL modules (and supermodules) which are assembled and tested at Rome and CERN prior to the final testing of each supermodule in the H4 test-beam. Much of the information collected during this phase will be needed not only to construct the detector but also for its calibration, to allow accurate simulation of its performance and to assist in its maintenance over the lifetime of CMS.

The CRISTAL system is being developed to facilitate the capture of this physics data. It is being built to be compatible with EDMS (CERN's Product Data Management system), it uses an OODBMS (currently Objectivity) as the repository for the construction and assembly data and incorporates CORBA technology as the mechanism for distributing functions. CRISTAL will, in its first instantiation, monitor and control the production and assembly process of the ECAL Barrel lead tungstate crystals and their associated electronics. Each of the Barrel ECAL detector components will have its physical characteristics individually measured and recorded to ensure consistency of the crystal production process. Quality control must be enforced at each step in the fabrication process.

The CRISTAL system, consequently, supports the testing of detector parts, the archival of accumulated information, controlled access to data and the on-line control and monitoring of all so-called Production and Local Centres. The software employs workflow and product data management techniques, will be generic in design and will be reusable for other CMS detector groups. Furthermore, the CRISTAL software could potentially be used in any large-scale production and assembly environment in HEP including LHC accelerator or experiment construction. More detail on CRISTAL specifics can be found in [LeG96, LeG97, McC97].

A legacy Engineering Data Management System (called EDMS [Ham96]) is being used at CERN to manage, store and control all the information relevant for the conception, construction and exploitation of the LHC accelerator and experiments during their life cycle. EDMS defines these systems in terms of product and work breakdown structures (in a so-called "as designed" view). Figure 2 shows where EDMS and CRISTAL are used in a detector's lifecycle. Essentially EDMS is for the management of documents (such as CAD/CAM 3-D drawings) produced during the mechanical design phase of CMS development, whereas CRISTAL is used for physics data collection during the prototyping, construction, operation and maintenance phases of the project. Figure 3 shows the relation of CRISTAL with other important CMS software modules such as simulation programmes, calibration software and controls systems. Functionally, CRISTAL can read data from EDMS and for each defined product it can gather test data and physical characteristics for later physics use.

Engineering drawings, blueprints, construction procedures, planning and costbook information will be stored in EDMS. The workflow and part definitions required to control the ECAL construction process could, potentially, be extracted by CRISTAL from EDMS. CRISTAL uses definitional information to carry out the functions of production management and workflow management and tracks parts through the manufacturing, assembly and maintenance life cycle. The CRISTAL system provides the (so-called "as built") engineering view to information stored during ECAL construction and in doing so provides the physicist with the detector geometry. The detector geometry is built up using the planned detector layout and as a consequence of assembling, for example in ECAL, crystals into sub-units, sub-units into modules and modules into supermodules thereby establishing the position of one module with respect to another.

CRISTAL is also capable of providing support for other views onto the construction data. These views include:

- Calibration View. Where physicists will want to view and access part characteristic data, classified by readout channel rather than by detector composition, for experiment calibration and event reconstruction purposes.

- Maintenance View. Where engineers will refer to the production processes for assembly and disassembly procedures, update information collected through maintenance operations and design modifications throughout the experiments lifetime.

- Experiment Systems Management View. Where the "slow control" system can view the part production history for configuration and fault management purposes.

This paper considers how a CRISTAL system can be set up and used in practice to manage the construction of a CMS detector and how the data captured in CRISTAL can be accessed by physics applications adopting a calibration viewpoint.

# 2. Establishing a CRISTAL System

This section explains the purpose of setting up a CRISTAL system and the structures that must be established (in a database) for its successful operation. It identifies the environment in which a CRISTAL system should be used and introduces the roles of users interacting with its software, which are described in detail in the following section.

CMS assembly will be carried out by groups with responsibilities for sub-detector construction (or sub-sub-detector construction) such as the ECAL, Tracker, HCAL, ECAL-Barrel etc. Each group needs to be only loosely coupled to others for final detector integration and must preserve their autonomy during the assembly process. Each CRISTAL system is set up to manage the accumulation of physical characteristic data during the construction of a particular CMS detector. Each construction follows a specific production plan and each detector is assembled and tested in a step-wise fashion. Therefore, at any point in the overall assembly of CMS there could be several CRISTAL systems active: one for each logically separated assembly process. Clearly at installation time the scope of the CRISTAL system must be set e.g. ECAL-Barrel, ECAL, ECAL-HCAL, CMS. The scope of a CRISTAL system defines the extent to which the detector breakdown structure is used for defining CRISTAL products i.e. the root of a CRISTAL system provides the entry point of that system in the overall CMS assembly tree.

In addition to establishing the root or scope of the CRISTAL system it is necessary to declare each Centre that is federated into the system. The Central System (normally at CERN) is declared in addition to each Local Centre in which the CRISTAL software will run. The CRISTAL system uses a set of roles to define user access to its software and data and these are detailed in the following section. Each Local Centre will also have a set of Instruments defined in the database in terms of the commands it uses and the outcomes it expects following the execution of activities.

Figure 4 shows the minimum hardware configuration and setup for a small Local Centre. Clearly the number of Operator workstations required at each Local Centre is dictated by the number of concurrent users required to cope with the volume of parts handled at the Centre. Typically a small Centre may be configured with up to 5 Operator stations, while a Centre, such as CERN, dealing with large volumes of parts may have 10 or more Operator stations configured. Each Operator station requires a PC having at least a 200MHz processor, 64 Mbyte of memory and a Barcode reader. Each station needs to be equipped with a Barcode reader since individual parts, worked on concurrently by individual Operators located in different rooms in a Centre must be uniquely identified. Disk space on the Operator stations is not important, since all CRISTAL data will be stored on a server, but inexpensive PCs these days are normally shipped with at least 1 Gbyte of disk space. Each Local Centre must have a server on which the Objectivity database is located. This machine (either a large PC or a Unix device) requires at least a 300 Mhz processor, 128 Mbyte of memory and a 9 Gbyte hard disk. The Central System at CERN, where the complete set of data is captured in Objectivity for a detector and is securely backed up by CERN services, will be equipped with storage of up to 1 Terabyte of which 100Gbyte will be available online at any one time.

Figure 5 shows the software architecture of a Local Centre. CRISTAL software comprises a set of Instrument Agents (for handling Instrument-CRISTAL communication), a set of Product Managers (one for each part which saves all part-related data in the Objectivity database), a Local Centre Manager which supervises the data gathering, a set of Digital Control Panels (DCPs) which handle all user interaction with CRISTAL and a Data Duplication Manager which handles all duplication of data between the Local Centre and the Central System based at CERN (which itself provides the necessary back-up facilities for the data). All software modules, including user supplied software, is integrated in CRISTAL using a CORBA (Common Object Request Broker Architecture) facility called an ORB. Consequently, each Local Centre must be provided with the Orbix CORBA software from Iona Technologies as well as the Objectivity V5 database in which all product data is captured.

In the case of ECAL the assembly production plan is defined in steps from individual crystals, APDs and capsules, through sub-units and modules to supermodules. To enable management of this production procedure (and to facilitate essential pre-experiment activities) it is necessary to have knowledge of the detailed breakdown of the detector. Without this detector breakdown structure it would be impossible to keep track of the processes of detector component production and assembly, which is the main purpose of CRISTAL. Figure 6 shows an overall detector breakdown structure for CMS (i.e. each CMS detector and its principal components) and figure 7 shows the ECAL Barrel breakdown as it was designed including quantities of each part (e.g there are 36 Supermodules in the Barrel and 4 kinds of Modules in each Supermodule). The part quantities are used to allow visualisation of the complete Barrel structure and to provide mechanisms for the expansion of the type structure into an individual parts structure [LeG97]. It is this individual "as-built' parts structure which is ultimately required for gathering detector physical characteristics. In other words, the overall CMS structure of figure 6 can be zoomed to show the ECAL Barrel structure (figure 7) which can, in turn, reveal the detector breakdown structure for a sub-module of a specified type (figure 8). Each sub-module type comprises an alveolar structure type, a tablet type, five sub-units of type Left and five of type Right. Each sub-unit comprises a Crystal type and a Capsule type. Note the use of a generic type definition in the detector breakdown structure - each instantiation of the type has a set of physical characteristics which is ultimately stored in the ECAL production database for later use, as described later in this note.

In order to provide the mechanisms for controlling detector production, it is essential to specify an assembly sequence for each type of each element in the detector breakdown structure (e.g. Sub-Unit, Crystal, Capsule, SubModule) as shown in figure 8. An assembly sequence is simply an ordered sequence of activities, each of which may require resources that will take place at a certain location and may or may not be carried out in parallel with other activities. Resources that may be allocated are termed Agents and are of the type Instrument, Human or User-provided code. Agents execute activities in CRISTAL and are the initiators of the data gathering process. The assembly sequence, which is executed, may itself be composite in nature. For example the Crystal assembly may involve a characterisation step which itself can be broken down into longitudinal and transverse transmission measurements, 3D measurements and light yield measurements (as shown in figure 9).

Figure 10 shows that activities (in the case shown: 'glue capsule onto a crystal') act on products (crystal, capsule) if certain *start conditions* are fulfilled. The start conditions are the pre-requisites for the activity to take place and include (at least) the products involved in the activity. The conditions are tested at the outset of the activity and the activity can only progress when all start conditions have been fulfilled. Each activity has a description, takes place at a specified Local Centre(s) (e.g. Rome or CERN), can be allocated resources (e.g. an Operator Agent), can have a specified minimum and maximum duration, may optionally be repeatable and, when completed, produces an outcome which is tested against a set of so-called *end conditions*. All of this information is specified for a product type named in the central ECAL production database, for a particular class of activities, by the Coordinator. Each instance of the activity is actually run at the specified Local Centre and, on completion, if successful, physical characteristics are computed and stored against the tested product for future use in, for example, maintenance.

## 3. Roles in CRISTAL

In any CRISTAL system the Central System will have the following roles: Coordinator, Centre Supervisor, Physicist and Information Manager and each Local Centre will have the following roles: Centre Supervisor, Operator and Information Manager. It is through the use of roles that privileges on and access to the CRISTAL system is controlled. This section outlines the responsibilities for each of these user roles.

Operators will be located in the various local centres (production and assembly), will perform all the activities on the products and will operate all the instruments which provide the physical characteristics. Operators will register parts in, or make parts known to, the CRISTAL system. Prior to registration, no data can be saved in the ECAL production database for parts; following registration all part-data is associated with an unique identifier in CRISTAL. In addition, Operators will manually capture all the parts information not automatically collected by the instruments. Operators will be technicians with no detailed knowledge in computer technology. They are expected to use the CRISTAL system to capture the information relevant to the production process and to operate computer controlled instruments which will populate the data storage automatically. Up to 10 to 12 operators per centre are expected to use CRISTAL interactively. An operator will interact with the system to start/stop a session, to browse part characteristics and to execute tasks on parts.

There will be one active Centre Supervisor per centre (Production Centre, Local Centre and Central System). The Centre Supervisor will be responsible for the quality control of the products or detector parts locally produced or assembled. His main task will be to accept or reject the various parts and to monitor and optimise the detector assembly. Centre Supervisors will also be responsible for satisfying orders for products to be shipped between centres. Here one order corresponds to one physical shipment of a list of parts between one centre and another. The Centre Supervisor of the Central System will browse the engineering information system to supervise the process and will issue the part orders, which are then fulfilled by Centre Supervisors at the outlying Local Centres. Centre Supervisors will normally be engineers or physicists with a background in High Energy Physics and/or in scintillating material sciences. They will be expected to browse the engineering information system, to supervise the process and to interact with CRISTAL on a daily basis. Being the supervisor of a local centre, he will be able to execute any of the tasks that an operator can execute.

Physicists will be interested in obtaining information about a detector of the CMS experiment. They will perform two distinct activities. On the one hand, they will consult the ECAL production database using standard browsing facilities. On the other hand, they will write software programs to perform analysis tasks using the data contained in the ECAL production database. All physicist access will be performed on the Central System.

The Coordinator, located at the Central System, will be responsible for the full detector construction lifecycle.

He will be a physicist with a background in High Energy Physics and in scintillating material sciences. He is expected to provide the detector quality control technical. In particular, the Coordinator will be responsible for the definition of the production procedures, the workpackage steps and for the specifications of data capture operations related to the various measurements. The Coordinator will interact with the system, on installation, to define a local centre, to create production scheme elements (parts, characteristics, fields, tasks) to handle versions of the production scheme (a complete set of chained tasks which will control the production process), to handle requests of acceptance/rejection on parts and to browse the central storage to monitor production.

The Information Manager will be the user responsible for the maintenance of the CRISTAL Prototype 2 system in a centre. He will be a physicist or an engineer with a background in computing. His main responsibility will be to make sure that the system is installed and fully operational and available in the various centres. He will also verify that no information is lost when being migrated from one centre to another. In addition, in the Central System, this user will be in charge of the maintenance of the ECAL production database. The Information Manager will interact with the system to establish a CRISTAL system, to set-up a local centre, to manage local storage and to handle system upgrades.

## 4. Running a CRISTAL System

As mentioned earlier it is the Operator in each Local Centre who will be responsible for the day-to-day running of the activities which are tracked in the ECAL production database. The Operator will browse the data and the outcome of activities already accumulated in CRISTAL and will execute activities following a predefined sequence (or production scheme). This sequence will have been defined centrally and will have been supplied to each Local Centre by the Coordinator and ultimately will have been made available as the current scheme by the local Centre Supervisor. Activity execution is controlled through the use of start and end conditions. An example of the Operator screen interface is included in figure 11.

Activities are initiated by Operators once the barcode of a particular product has been swiped – barcode swiping is the mechanism by which products are securely and uniquely identified with a CRISTAL identifier and allocated to an activity. Swiping the barcode also ensures that the parts on which an activity is to be performed (e.g. gluing capsule onto a crystal) are physically located in the centre at which the Operator (and, in the example, a gluing device) is located. Following execution of the activity, individual part data is associated with individually distinguishable parts, whereas data common between parts is attached to the element in which these parts are constituents, provided that the individual parts and their composite are uniquely identified in CRISTAL. For example, consider accessing the physical properties of an ECAL capsule with 2 APDs. Data related to an individual APD will be attached to the unique identifier of that APD, whereas data related to the behaviour of the two APDs (such as their average gain) will be attached to the capsule. If the individual parts are not uniquely identified in CRISTAL then all the data associated with the parts (e.g APDs) are attached to the next highest level at which the parts are uniquely identified (in this case, the capsule).

Quality control of the production process is maintained, once production is underway, through the use of each activity's end conditions. If the end conditions for an activity are satisfied the product is marked as 'conformant'; otherwise the product is marked as 'non-conformant'. Most products should normally be conformant and non-conformant products are referred to a Centre Supervisor for a decision on their further use. A specific exceptional procedure can be initiated by the Supervisor to attempt to reinstate a non-conformant product in the production line if deemed appropriate to do so. Products can be rejected by the Centre Supervisor and are then no longer used for CMS assembly. CRISTAL maintains a list of the conformancy status of all products (Conformant, NonConformant, Rejected) and the reason why they were marked as such.

As noted earlier, the overall CMS detector assembly and test procedure is distributed in nature; for the ECAL Barrel this distribution involves activity execution across numerous production and assembly Centres (in France, China, Russia, Italy and CERN). Management information regarding the configuration and operational state of the production is defined and stored centrally at CERN in a *production database* by the overall ECAL Coordinator. This object based repository also stores the definitions of the parts (or components) that make up the detector together with the definitions of the instruments used to produce or take measurements of parts. It stores descriptions of the lifecycle of each part and descriptions of the tasks and activities that must be performed on the part. Together this information constitutes a so-called *production scheme* which determines the order of tasks that can be applied to any part. The Coordinator issues versions of the production scheme for execution at the outlying production and assembly centres and to enable tracking of the overall production process (see figure 12). All the details of all the activities that have taken place and the associated physical characteristics that are consequently accumulated and the instruments involved in the measuring are eventually stored in the central repository at CERN (and are backed-up by the normal CERN procedures).

The data handling aspects of CRISTAL must be transparent to any users of the system irrespective of their location. Part, task, production scheme and system configuration data will be defined by the Coordinator at CERN thereby controlling the production in all Centres. Once the production centres are configured, parts will be produced and then shipped between local centres for testing and assembly. The Centre Supervisor of the Central System controls the flow of products between Centres by issuing Part Orders against which the shipping of parts between Centres is carried out (see figure 12).

Periodically, data transfer will take place between Local Centres and the Central System at CERN – normally overnight, on the movement of Parts between Centres or on initiation of a copy by the system supervisors. Each centre can, however, operate autonomously: it can continue to carry out tests on detector components and accumulate resulting data irrespective of whether it is connected to the central storage at CERN. On resumption of connection to CERN the Centre can upload its data and/or receive a new production scheme version or orders to/from the Central System. This process of data duplication from a Local Centre to the Central System or from the Central System to Local Centres is invisible to Local Centre users. Following the supply of a new production scheme version from the Central System, the local Centre Supervisor will ensure that the local instrument software has been appropriately upgraded prior to applying the new production scheme version.

## 5. Coping with Changes once Production has Begun

An important aspect of CRISTAL is that it is able to cope with user-defined updates in both the definition of products (e.g. an amendment to a type of detector component) and in the definition of activities to be performed on products (e.g. add a new component test, or amend an existing task). This is necessary since the activities and products are likely to evolve over the extended period of detector testing and assembly (1999-2005). This section briefly considers several examples of changes during detector assembly and details how CRISTAL copes with such updates. There is no detailed discussion of internal CRISTAL functionality in this section, instead an overview is presented of the philosophy adopted for handling changes in CRISTAL.

Production scheme changes can take place at three levels: at the Agent level, the activity level and at the product definition (or detector breakdown structure) level. Each level is now discussed. Firstly, over time, updates are likely in the Agents defined in the system. That is, in the commands used to drive measuring Instruments, in User-defined Code or in the allocation of roles (e.g. Operator) to activities. All updates in the production scheme are performed centrally (by the Coordinator) and are supplied to Local Centres following specification. These changes (typically a change in a data format of an activity outcome or a change in the definition of an Instrument command) are applied by the local Centre Supervisor at each Local Centre as soon as any production scheme upgrade has been sanctioned.

Activity updates are more difficult to schedule since there may be multiple currently active instances of the activity which is to undergo an update. CRISTAL handles this by determining a *change area* (or the scope of the update) in the current production scheme. If the product is outside of this change area the update is carried out as soon as the new production scheme has been applied by the Centre Supervisor. If, however, the product is undergoing an activity inside the change area, then the update is only allowed for that product once the change area has been traversed (see figure 13).

Changes to the detector breakdown structure are also handled by CRISTAL. Two examples indicate how they are catered for. In the first example consider a simple update to a product such as the inclusion of a temperature sensor to an existing product. As far as the definition of the product is concerned there is no real change and the update can be made for the 'new' product regardless of how many activities have already been completed for that product. In a more complex example consider the redefinition of a product (causing a potential change in detector geometry). Here the update can only be made to new instances of the product entering the production scheme (i.e at the start of their activities). Any existing products currently in the production scheme must follow that version of the scheme to completion with the 'old' definitions to ensure consistency. CRISTAL follows such a strategy with regard to complex detector breakdown changes.

## 6. CRISTAL as the Source of Precalibration Data

The ECAL production database will be first released in the summer of 1998 enabling ECAL physicists to collect all of the salient characteristics of each ECAL component. Once the detector has been tested and assembled it is essential for event recognition programs to have access to detector characteristics in order for event reconstruction and calibration to take place. The physics data collected during the detector assembly phase must be arranged and processed to allow population of a calibration database. This section details how the ECAL production database can be used as the source of calibration data.

For the purposes of physics analysis, CMS ECAL needs to produce a pre-calibration database which will hold all the physical parameters necessary for calculating the ECAL calibration static constants for a single ECAL supermodule. Production and calibration are two different views of ECAL detector data. Consequently, physicists must be able to extract *sets* of data relevant for determining calibration coefficients from the ECAL production database (figure 14) into this pre-calibration database. The information required for calibration includes data for crystals, APDs, supermodules and electronics. Crystal-specific information will include light yields, attenuation lengths, longitudinal and transverse transmission (as functions of wavelength) and any crystal non-uniformities. APD information needed for calibration includes quantum efficiency (vs. wavelength), gain and dark current vs. high voltage and nominal biases. All of this information will be captured and recorded in the ECAL production database as the detector is constructed step-wise from individual crystals through sub-units to super-modules.

The step-wise construction procedure leads to a data organisation necessarily different to that required for calibration - the structure of the production database follows the assembly ordering of the calorimeter, while the structure of the calibration database must follow the ECAL readout structure where the unit of detector that is normally considered is the readout channel. Calibration constants need to be determined for each readout channel (e.g. a crystal plus its APDs, electronics (ADC etc.) and fibres ) so that ADC counts can be translated into energy deposited in a single readout channel. Therefore, the calibration system must be able to *extract* subsets of physics data from the ECAL production database for the calibration of particular physics elements (or sets of detector components) even if these elements are specified in a manner orthogonal to that in which structures are defined in the ECAL production database.

All parameters needed for the calibration of a single supermodule will be extracted from the ECAL production database and a matrix {i,j} of readout channels (or so-called *physics elements)* will be built in the pre-calibration DB. To these data, information from test beam slow controls, data acquisition and monitoring is added to build a complete picture of the conditions under which calibration data was taken and, on completion of the calibration runs, these data are copied back into the production database (se figure 14). The production database then accumulates this calibration data for each supermodule and acts as the source of calibration data for all 36 supermodules. The final static calibration constants for the complete ECAL are extracted from the production database into a calibration database. Fast and efficient access from physicist programmes is allowed both to the pre-calibration database for a single supermodule in the H4 test beam and to the set of constants in the full calibration database (see figure 14).

To facilitate the implementation of navigation and data extraction tools, the ECAL production database will be object-oriented in nature and will be designed around a so-called *meta-model* [BLeG97]. This design will provide isolation of any modification of the database content from any software accessing it. The extraction facilities will be designed sufficiently generically to be used as the basis for ad-hoc database queries by physicists for the purposes of subsequent ECAL physics analysis as discussed in the next section. The design and development of the navigation and extraction utilities will be carried out by members of the CRISTAL team and members of the monitoring and pre-calibration physicists of IPNL, LPNHE and CEA/DAPNIA.

## 7. Generalising Physicist Access to CRISTAL Data

The calculation of calibration coefficients for particular readout channels is just one example where the ECAL production database needs to be consulted for physics data gathered during assembly and testing. In this example the 'physics element' of interest is a readout channel which is the basic unit of the calibration system. Other ECAL applications will need to extract sets of data from the production database e.g. for control and monitoring of the equipment, for alignment and geometry and, ultimately, for physics reconstruction programmes and, in each case, sets of physics elements can be specified for that purpose. For each, the application software will need to traverse the ECAL production database and to extract data, for the physics elements specified by a physicist, into an application-specific repository even when the physics elements do not necessarily follow the structures defined in the production database. In the next paragraph the mechanisms for providing this navigation and extraction are outlined.

The ECAL production database has been designed using object-oriented techniques with a view to providing reusability, flexibility and interoperability and to aid in the extraction process. An approach has been followed in the design of the CRISTAL data model which ensures that the database components are *self-describing*. That is, the objects retain knowledge both about their dynamic structure and how objects are related and this knowledge is available for applications to interrogate and to use as a basis of navigating the CRISTAL database. This design approach leads to so-called *data independence* in CRISTAL i.e. a separation of the physics data from any client applications which access it, so that any changes in the specification of, for example, physics elements does not require recompilation of the accessing software. The application software (e.g. the calibration system, or physics analysis programmes) does not need to know the relationships between the database objects but rather inquires of those objects how they should interact with them to satisfy a particular purpose, for instance, to extract calibration information. This approach, sometimes called a meta-modelling approach [BLeG97], allows the capture of knowledge alongside the objects, enriching them and facilitating self-description and data independence. Meta-modelling will assist the extraction of data for physics elements provided a *meta-query* mechanism is developed which can navigate the database, can interpret the structures in the production database and can present the data in a form meaningful to the physicist. Meta-queries could be used to assist, for example, in finding data that matched user criteria ("pattern-matching") in calibration or analysis programmes.

In conclusion it is the definition of detector components (using Product Data Management or PDM software), the tracking of each of the assembly sequence activities (using workflow management software) and the gathering of physics information at stages in the execution of the assembly activities that are the main goals of the CRISTAL monitoring system. Having captured this data the ECAL production database can be used as the information source for physics analyses, for example, in the gathering, management and presentation of data specific to calibration runs. Access to the data resident in the ECAL production database can be made general by the provision of a meta-query facility which can browse the data structures in the database and present data in a format meaningful to the physicist.

## Acknowledgments

The authors take this opportunity to acknowledge the support of their home institutes. In particular, the support of P Lecoq, J-L Faure and M. Pimia is greatly appreciated. Richard McClatchey is assisted by CERN and by the Royal Academy of Engineering under their Technology Foresight Award scheme during 1998 and acknowledges their support.

**Figures**

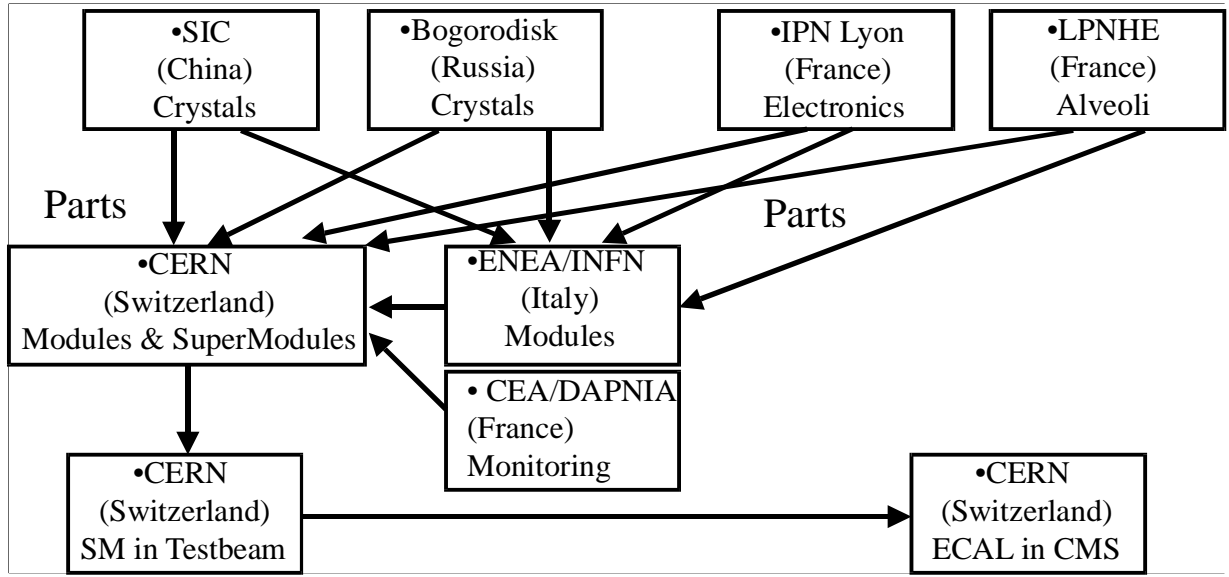

Figure 1: The distributed construction of CMS ECAL Barrel

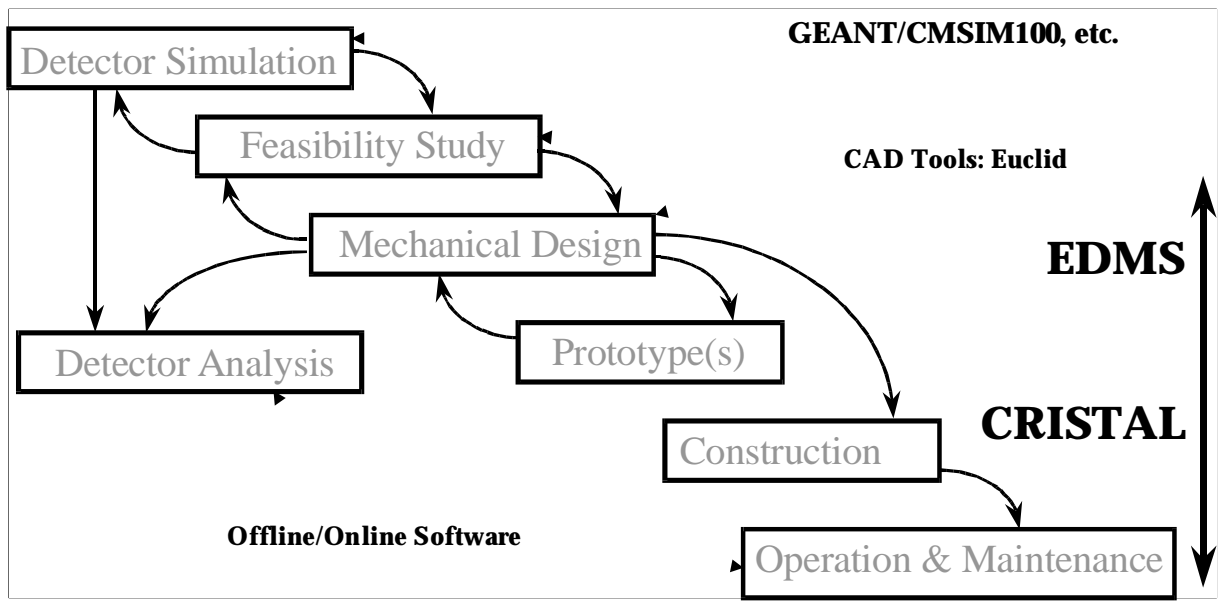

Figure 2: The role of EDMS and CRISTAL in the detector construction lifecycle.

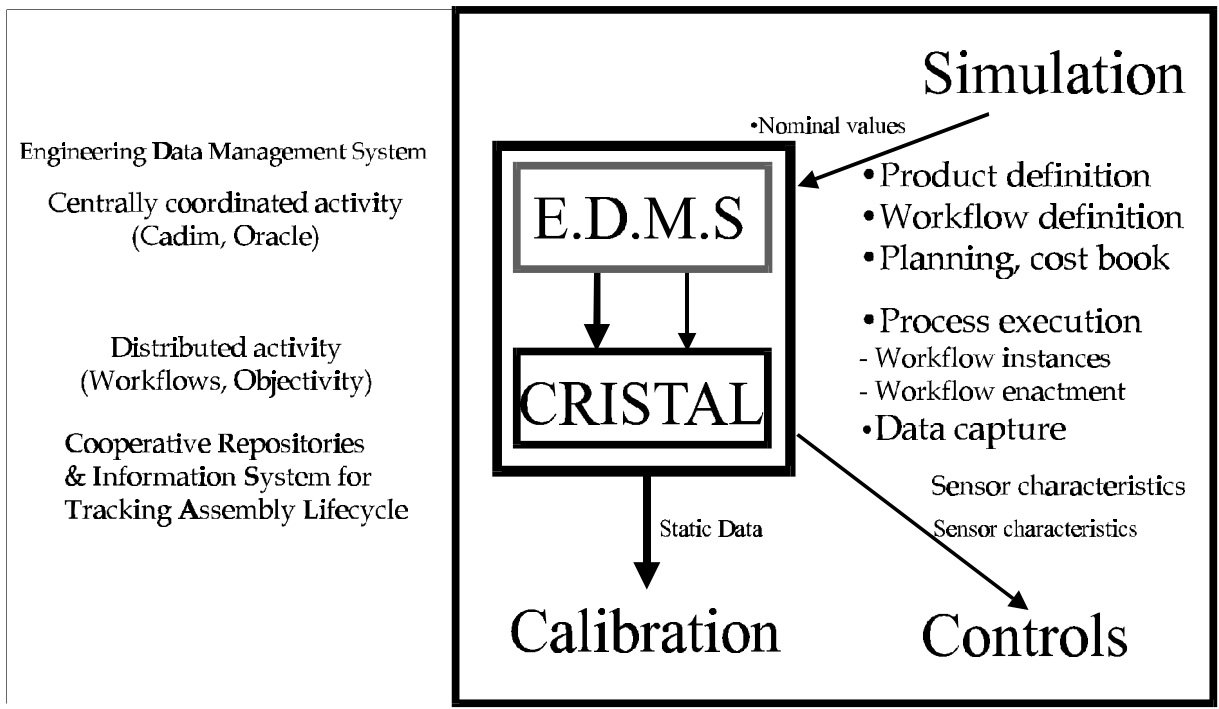

Figure 3: The interaction between CRISTAL and other CMS software.

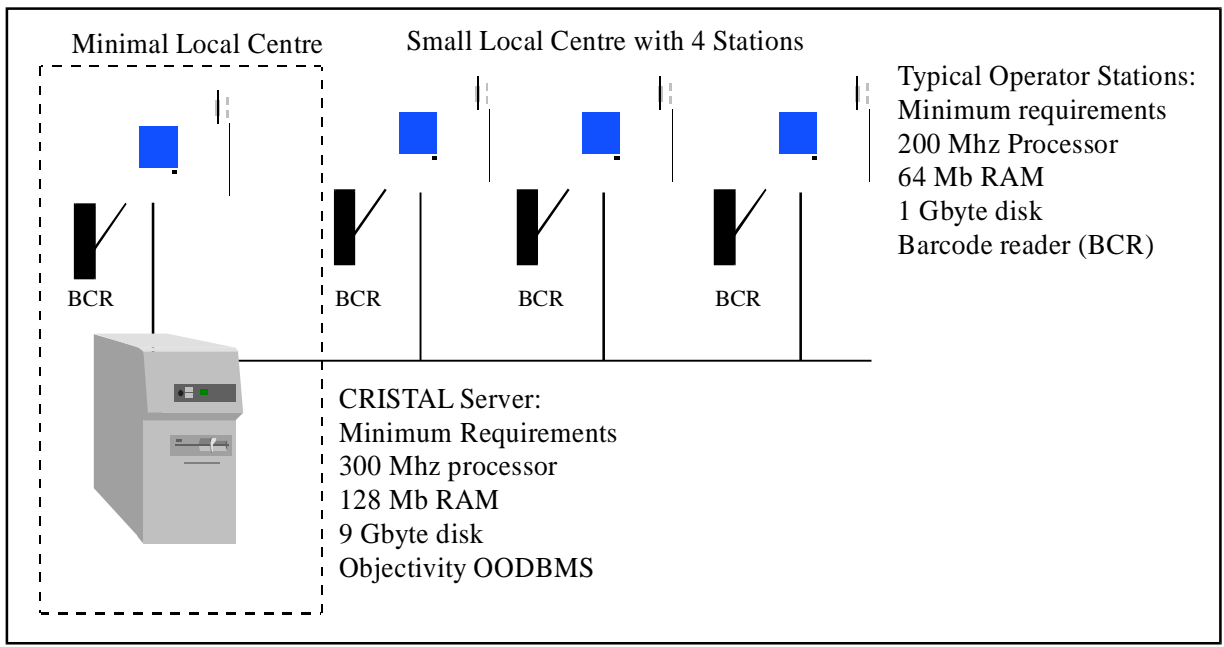

Figure 4: Local Centre minimum hardware requirements

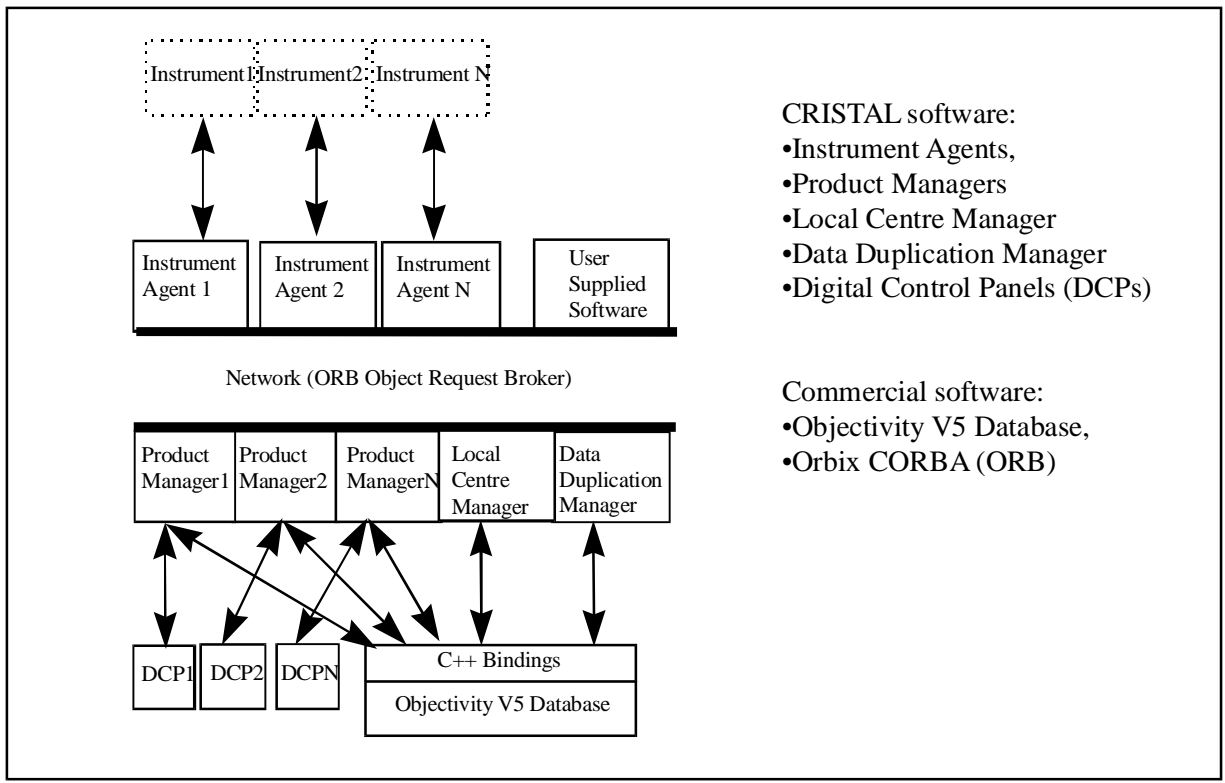

Figure 5: Local Centre software architecture

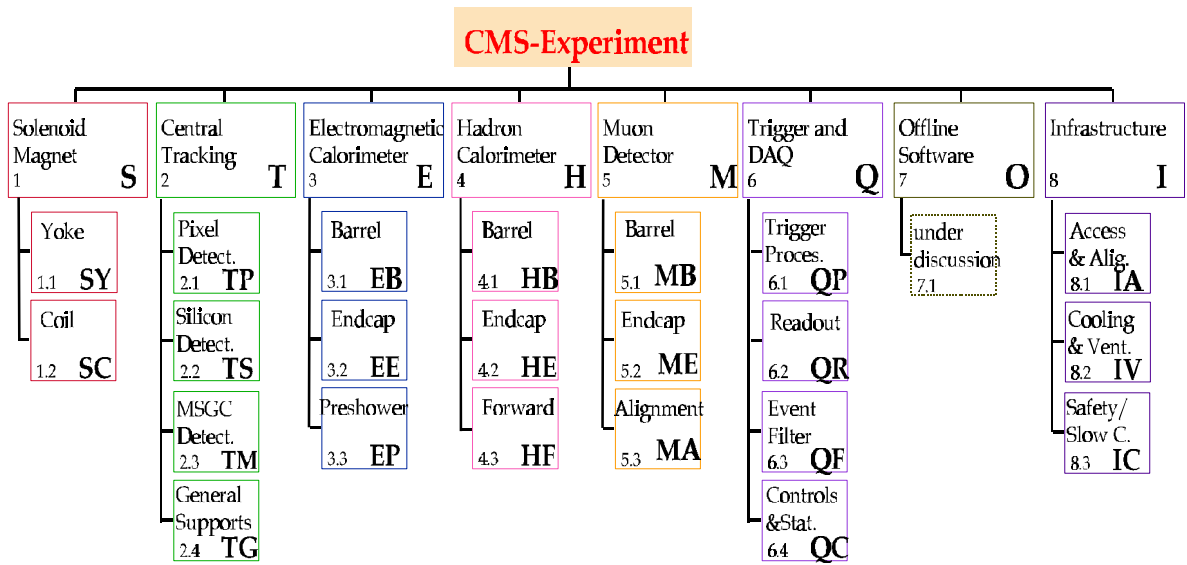

Figure 6: An overview of the complete CMS detector breakdown structure

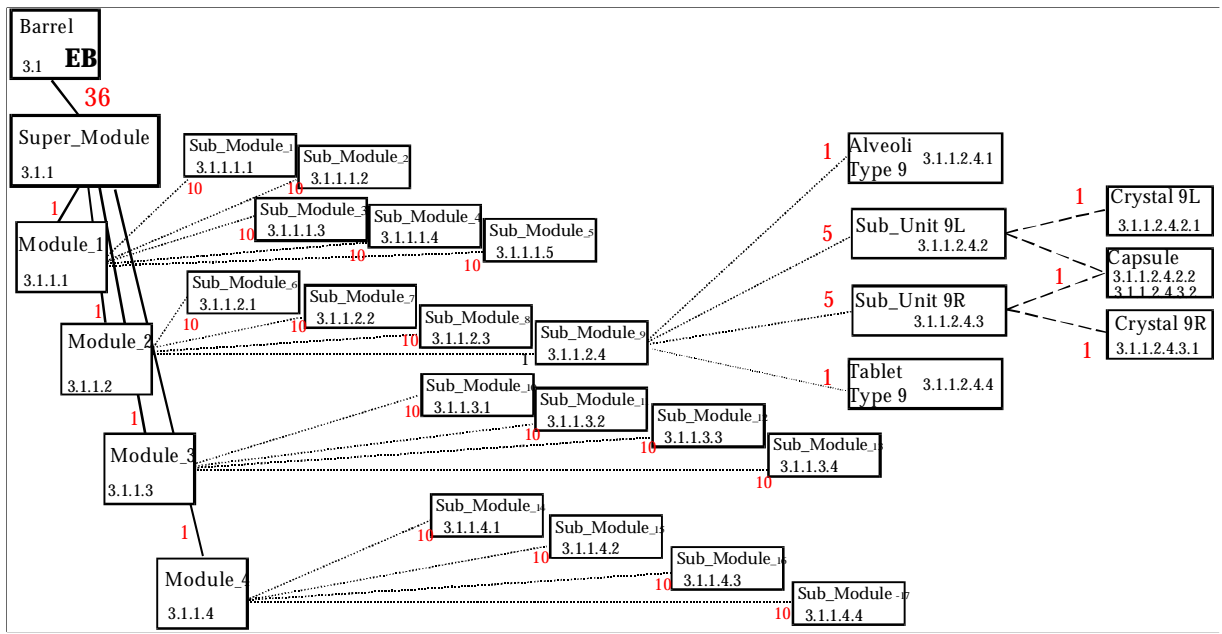

Figure 7: The ECAL Barrel breakdown structure

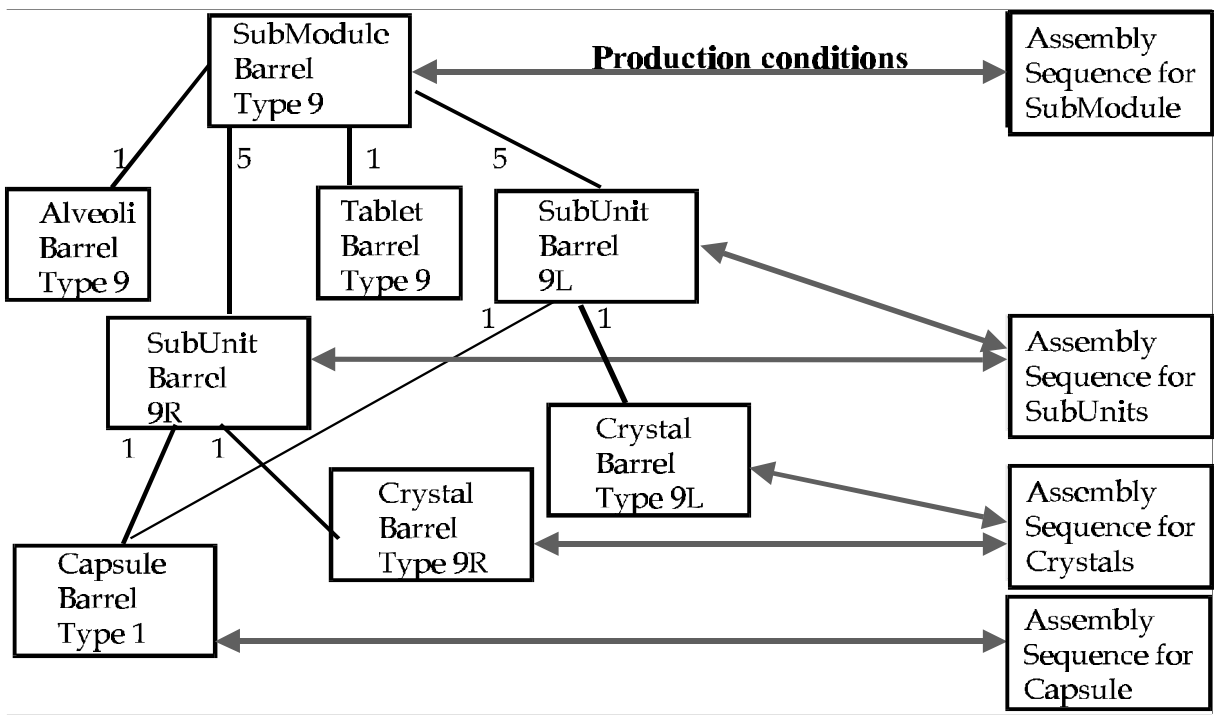

Figure 8: An ECAL Barrel Submodule structure and associated assembly sequences

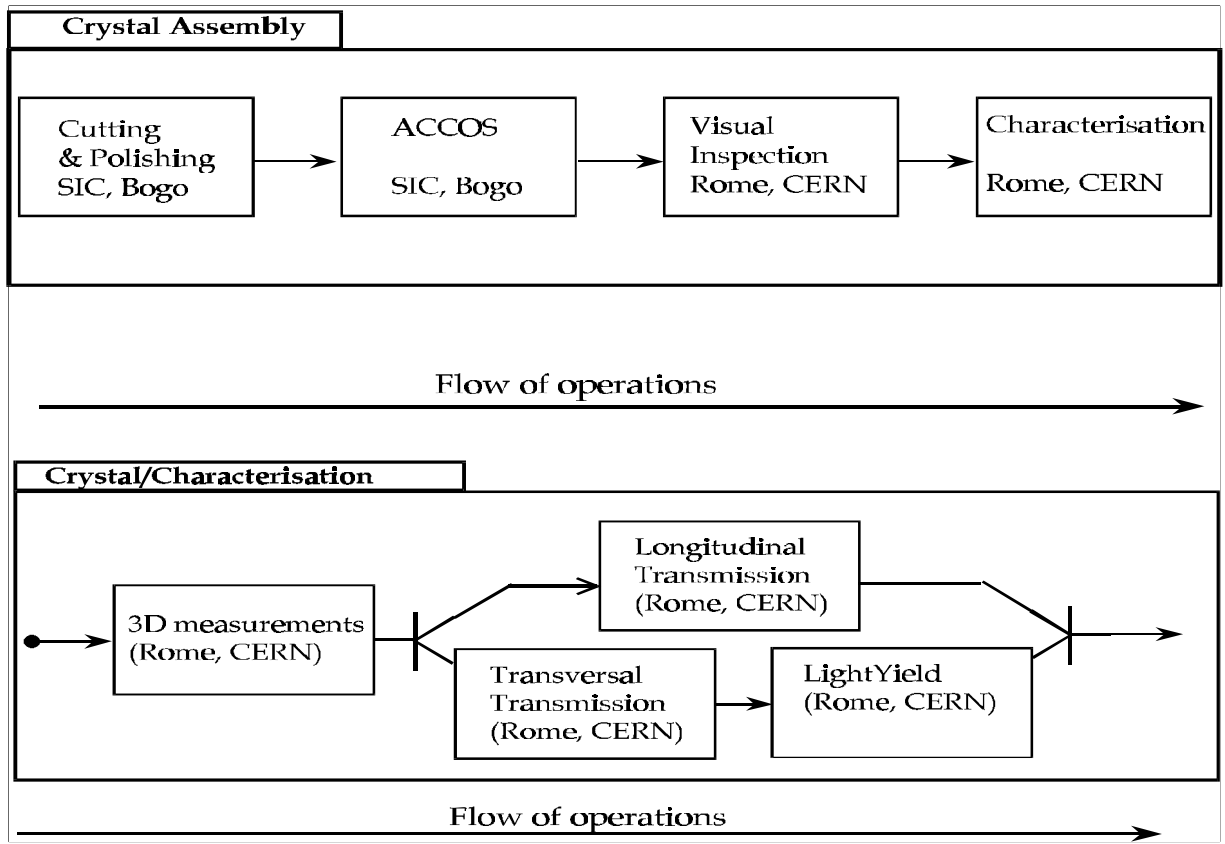

Figure 9: Examples of composite and elementary assembly sequences in CRISTAL.

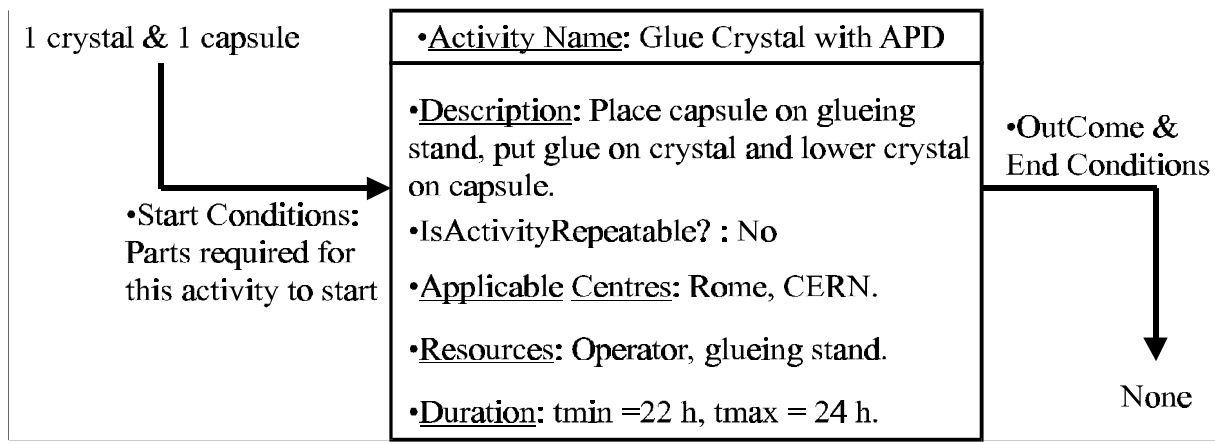

Figure 10: An activity definition showing resources, duration and production conditions.

Figure 11: An example of the CRISTAL operator interface.

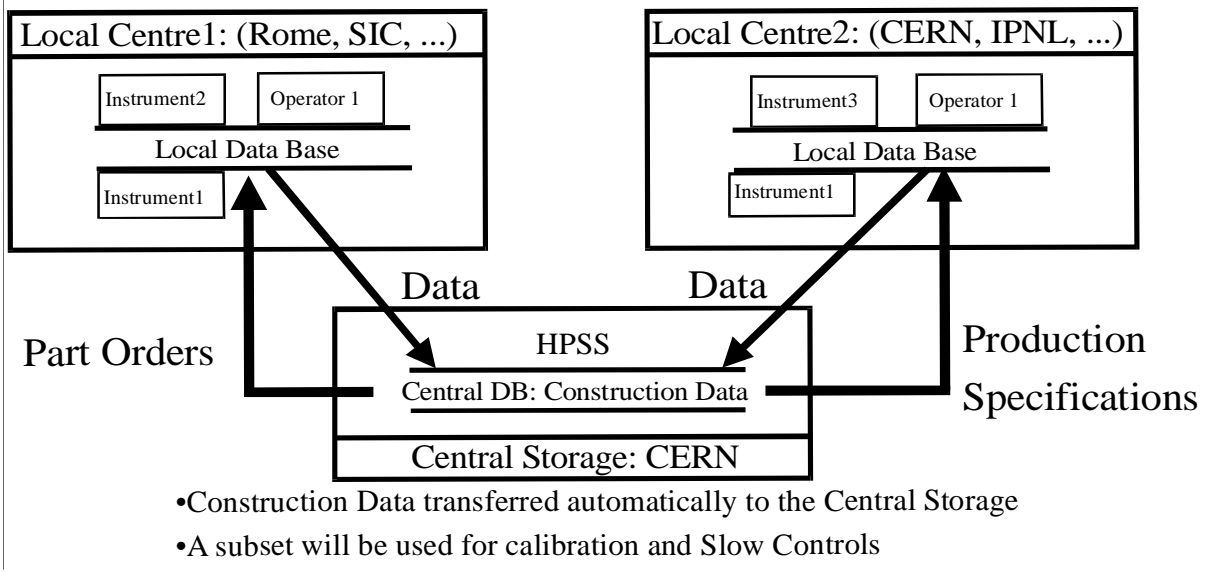

Figure 12: CRISTAL distributed production : movement of data, orders and specifications between Centres.

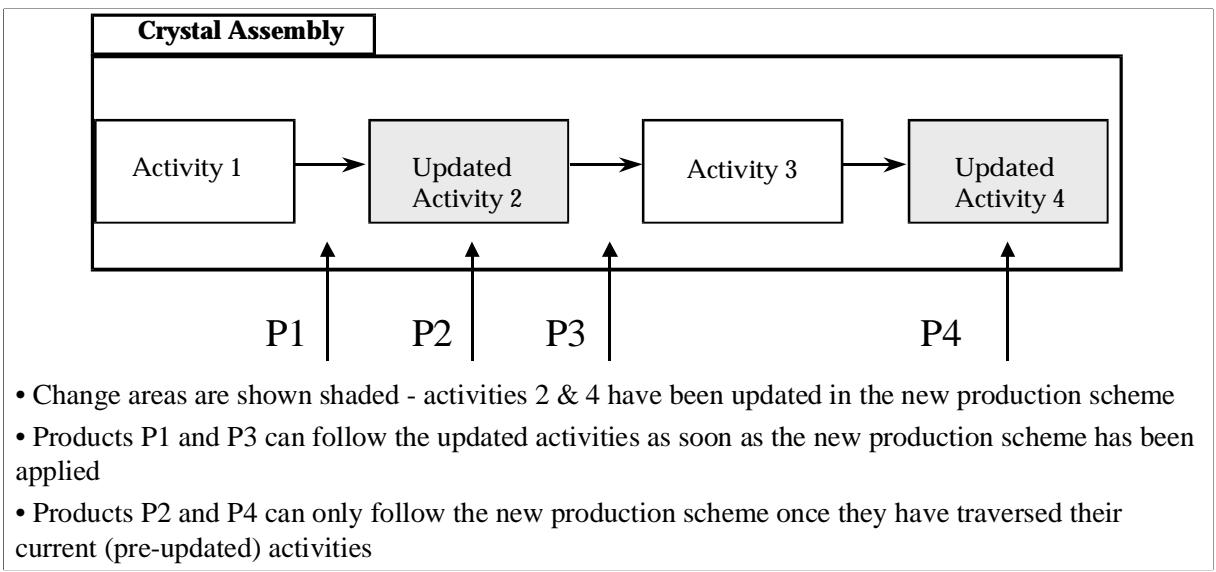

- Change areas are shown shaded - activities 2 & 4 have been updated in the new production scheme
- Products P1 and P3 can follow the updated activities as soon as the new production scheme has been applied
- Products P2 and P4 can only follow the new production scheme once they have traversed their current (pre-updated) activities

Figure 13: Handling dynamic changes in activities.

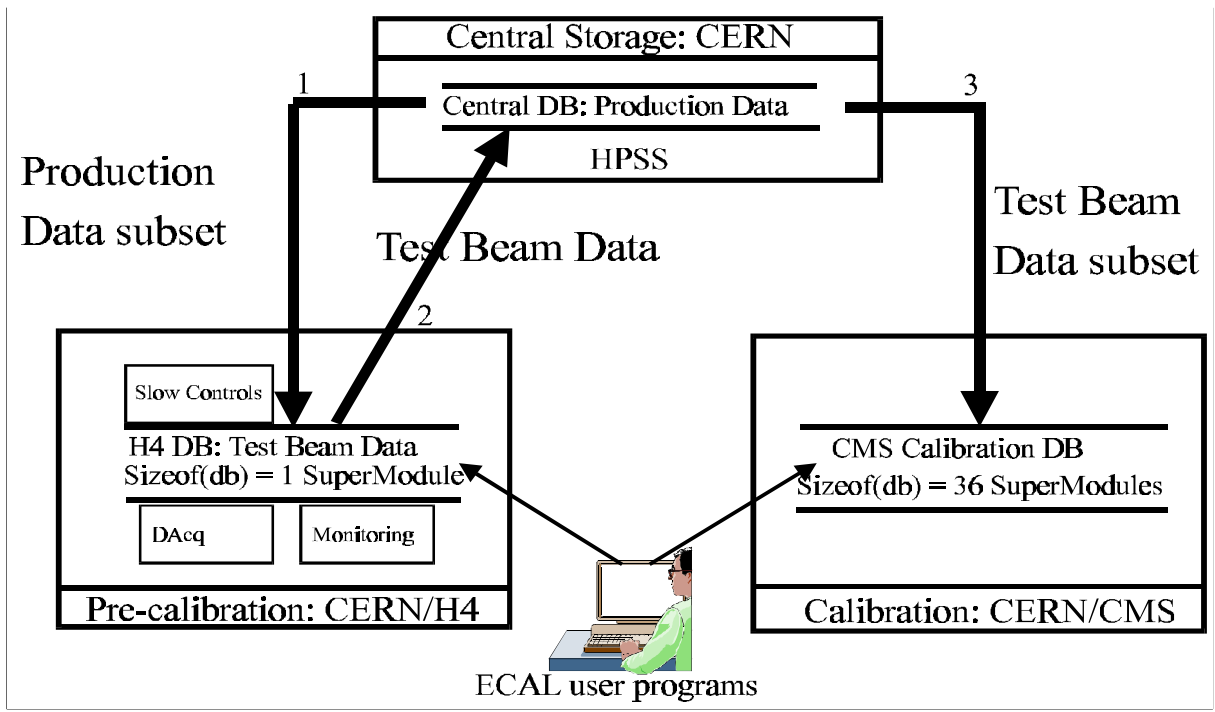

Figure 14: The use of CRISTAL data in (pre-)calibration databases.